\documentclass[final, natbib]{svjour3}
\usepackage{rotating}
\usepackage[T1]{fontenc}
\usepackage[utf8]{inputenc}
\usepackage{newtxtext,newtxmath}
\usepackage{url}
\usepackage{hyperref}

\usepackage{etoolbox}
\newcommand*{\affaddr}[1]{\textsuperscript{#1}} 
\newcommand*{\affmark}[1][*]{\textsuperscript{#1}}

\usepackage{bm}
\usepackage{soul}
\usepackage{xcolor}%
\usepackage[T1]{fontenc}

\makeatletter
\journalname{Journal of Low Temperature Physics}

\bibpunct{}{}{,}{s}{}{,}

\begin{document}

\newcommand{\hdblarrow}{H\makebox[0.9ex][l]{$\downdownarrows$}-}
\title{High-sensitivity Kinetic Inductance Detector Arrays for the PRobe far-Infrared Mission for Astrophysics}
\titlerunning{High-sensitivity KID Arrays for PRIMA}
\authorrunning{L. Foote et al.}

\author{L. Foote\protect\affmark[1,2] \and C. Albert\protect\affmark[1, 2] \and J. Baselmans\protect\affmark[3] \and A. D. Beyer\protect\affmark[2] \and N. F. Cothard\protect\affmark[4] \and P. K. Day\protect\affmark[2] \and S. Hailey-Dunsheath\protect\affmark[1] \and P. M. Echternach\protect\affmark[2] \and R. M. J. Janssen\protect\affmark[1, 2] \and E. Kane\protect\affmark[1, 2] \and H. Leduc\protect\affmark[2] \and L.-J. Liu\protect\affmark[1] \and H. Nguyen\protect\affmark[1, 2] \and J. Perido\protect\affmark[5] \and J. Glenn\protect\affmark[4] \and J. Zmuidzinas\protect\affmark[1, 2] \and C. M. Bradford\protect\affmark[1, 2]}

\institute{\affaddr{1} California Institute of Technology, 1200 E California Blvd, Pasadena, 91125, California, USA. Email: lfoote@caltech.edu\\
\affaddr{2} Jet Propulsion Laboratory, California Institute of Technology, 4800 Oak Grove Dr, Pasadena, 91109, California, USA \\
\affaddr{3} SRON, Netherlands Institute for Space Research, Niels Bohrweg 4, 2333 CA Leiden, Netherlands \\
\affaddr{4} NASA Goddard Space Flight Center, 8800 Greenbelt Road, Greenbelt, 20771, Maryland, USA \\
\affaddr{5} University of Colorado Boulder, Boulder, 80309, Colorado, USA \\
}

\maketitle

\begin{abstract}

Far-infrared (far-IR) astrophysics missions featuring actively cooled telescopes will offer orders of magnitude observing speed improvement at wavelengths where galaxies and forming planetary systems emit most of their light. The PRobe far-Infrared Mission for Astrophysics (PRIMA), which is currently under study, emphasizes low and moderate resolution spectroscopy throughout the far-IR. Full utilization of PRIMA’s cold telescope requires far-IR detector arrays with per-pixel noise equivalent powers (NEPs) at or below $1\times10^{-19}\;\mathrm{W}/\sqrt{\mathrm{Hz}}$. We are developing low-volume aluminum kinetic inductance detector (KID) arrays to reach these sensitivities. We describe the development of our long-wavelength ({80-265 $\mu\mathrm{m}$}) array approach and present multitone measurements of a 1,008-pixel arrays. We measure an NEP below $1\times10^{-19}\;\mathrm{W}/\sqrt{\mathrm{Hz}}$ for 73\% of the measured pixels.
\end{abstract}

\section{Introduction}
    The far-infrared (far-IR) offers a powerful window into a range of astrophysical processes.  Far-IR light flows unimpeded through dusty regions in our galaxy and across cosmic time, tracing the processes and contents which are hidden to ultraviolet and optical measurements. Far-IR spectroscopy and spectrophotometry are excellent techniques for studying the history of star formation and black hole growth in the Universe, and revealing fundamental properties of planet-forming disks. Though rich scientifically, the far-IR has lagged its counterparts at optical, near-IR, and millimeter wavelengths because it requires observing from space, and ideally, a cold telescope and detectors limited by radiation from the astrophysical background (zodiacal light). The international community has pursued this for some time with the SPICA \cite{nakagawa_next-generation_2009, roelfsema_spicalarge_2018} and Origins \cite{meixner_origins_nodate-1} concepts, and is now considering options for a NASA-led far-IR Probe.  One such concept is PRIMA, the PRobe far-Infrared Mission for Astrophysics \footnote{https://prima.ipac.caltech.edu/}.  With suitably sensitive detectors, PRIMA will achieve a two-order-of-magnitude improvement in sensitivity (four orders of magnitude in observing speed) over previous far-IR missions. 
    \par We are developing large-format arrays of low-volume kinetic inductance detectors (KIDs) to satisfy the requirements of PRIMA’s spectrometer, the Far-InfraRed Enhanced Survey Spectrometer (FIRESS). FIRESS is a grating spectrometer that targets the 25 to {235} $\mathrm{\mu m}$ wavelength range and provides a spectral resolution in low-resolution mode of {$R=85-150$} and a tuneable spectral resolution in high-resolution mode, with a maximum of $R=17,000$ at $25\;\mu\mathrm{m}$ and $R=4,400$ at $112\;\mu\mathrm{m}$. Each of FIRESS’ four spectrometer modules will contain two arrays of 1,008 pixels. All arrays will have the same pitch and format, but with custom-designed inductor / absorbers {and lenslets} optimized for its respective waveband. The instrument thus has 8,064 pixel in total, each of which needs to satisfy a noise-equivalent power (NEP) of $\approx 1\times10^{-19}\;\text{W}/\sqrt{\text{Hz}}$, a number that is approximately equal to the natural photon background noise in a FIRESS resolution element.
    \par Pioneering work at SRON demonstrated NEPs  below $1\times10^{-19}\;\text{W}/\sqrt{\text{Hz}}$ using a lens-antenna coupled detector with a low-volume aluminum active element \cite{baselmans_ultra-sensitive_2022, bueno_ultrasensitive_2018}. {We are pursuing KIDs based on multi-mode absorbers patterned in aluminum, also coupled via lenses. These absorbers offer a somewhat more versatile coupling option, particularly for short wavelengths, and are a good match to the instrument architecture for PRIMA. We are developing demonstration arrays at two wavelengths: {25-80 and 80-265 $\mu\mathrm{m}$}. Both arrays use} Nb plugs to limit quasiparticle diffusion, which increases the effective quasiparticle lifetime. Both arrays also achieve a sufficient level of cosmic ray mitigation for PRIMA \cite{kane_ltd2023}. The demonstration arrays also incorporate two different, but interchangeable, capacitor approaches. The short-wavelength arrays use Al/a-Si:H/Nb parallel plate capacitors (PPC) to reduce the pixel pitch. The short-wavelength array has shown good yield, optical efficiency, and fractional frequency noise but has lower-than expected response due to short quasiparticle lifetimes \cite{cothard_ltd2023}. The long-wavelength arrays use Nb interdigitated (IDC) capacitors, which have met our noise sensitivity requirements in previous devices. Single-tone NEP measurements of devices from an earlier prototype array than the array measured here are presented in Hailey-Dunsheath et al. \cite{shd_singletone_2023}.
    \par In this paper we present the first results of a flight-like KID array for PRIMA’s FIRESS spectrometer.  We focus on the initial characterization of a 1,008-pixel long-wavelength ({80-265 $\mu\mathrm{m}$}) array, which meets the PRIMA per-pixel sensitivity target, and is undergoing system integration with multiplexed readout for a demonstration of yield. The array layout is described in Sect.~\ref{sec:layout} and the electrical characterization and noise measurements using an RFSoC-based multitone system are presented in Sect.~\ref{sec:meas}. A summary of the results as well as an outlook for future improvements are presented in Sect.~\ref{sec:summary}.

\section{Long-wavelength array design}
\label{sec:layout}
    \begin{figure}[h]
        \centering
        \includegraphics[width=4.76in]{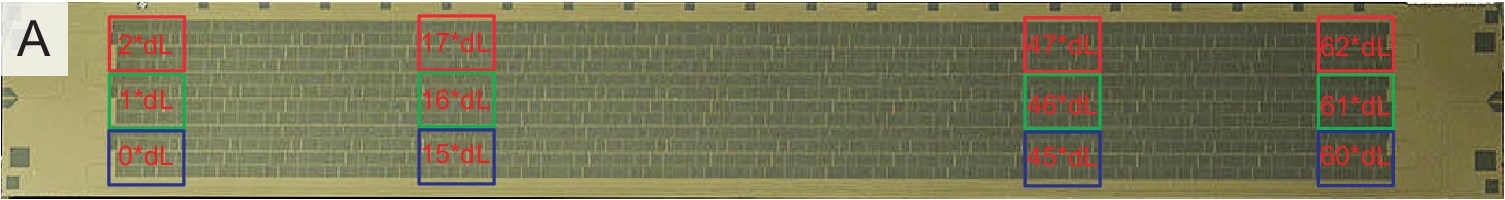}
        \includegraphics[width=4.76in]{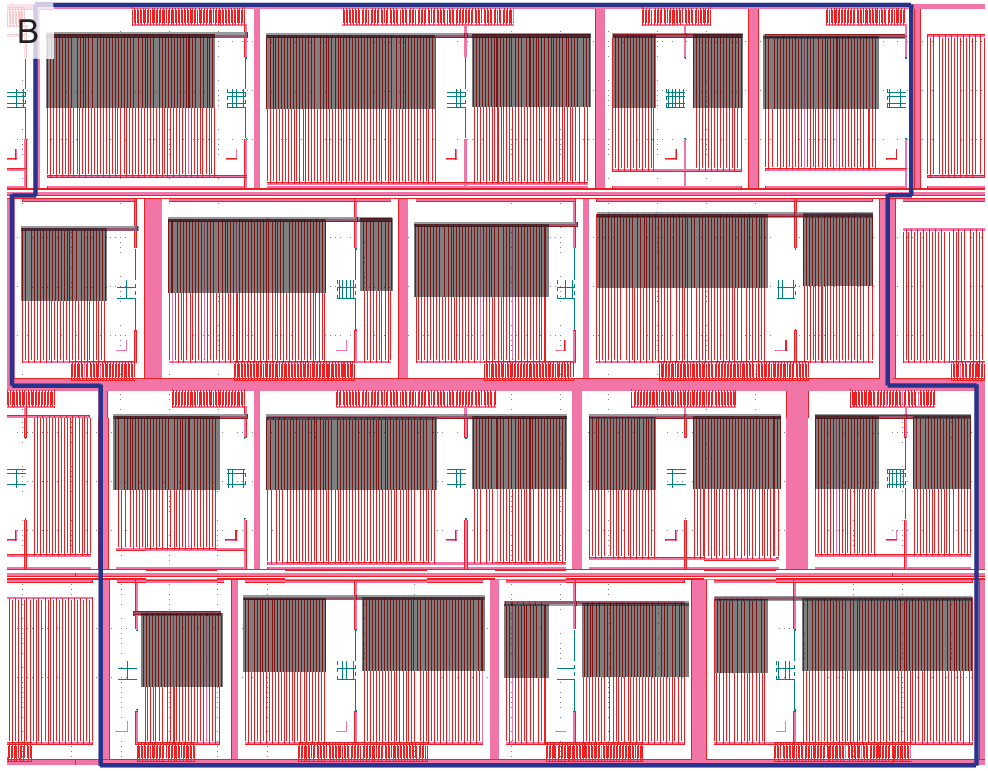}
        \caption{Array layout for a flight-like long-wavelength PRIMA FIRESS array. Panel (a) shows a photograph of the 12 x 84 pixel array. Individual pixels can be identified in the niobium (greenish yellow) as can the single coplanar waveguide (CPW) transmission line that snakes through the arrray and has larger bondpads in the middle of the left and right side edge. Around the edge, holes are left in the niobium to expose the Silicon wafer (black squares) and allow space for DC test structures (as seen above the left-most column of resonators). Panel (b) shows design of the 4 by 4 pixel unit cell (inside blue outline) from which the array is constructed. Each detector consists of an aluminum (green) absorber/inductor, one or two niobium (red) main IDC capacitors, IDC capacitor to ground, and coplanar plate capacitor. The latter couples the KID to the CPW transmission line, which runs below rows 1 and 3. The $\lrcorner$ shapes are alignment markers for the absorber patterning. The 16 KIDs in a unit cell are log-uniformly spaced within the design bandwidth by varying the number of capacitor tines (capacitor width). Resonator frequencies are further varied across the array by reducing the tine length by steps of $dL=0.7\ \mathrm{\mu m}$ for each instance of the unit cell, as indicated in panel (a). This is achieved by lithographically shifting the top of the main IDCs (shaded regions) downwards.}
        \label{fig:layout}
    \end{figure}
    \par In this paper we study the  12 by 84 pixel array shown in Fig.~\ref{fig:layout}. The individual pixels are based upon the high-sensitivity pixel design presented by Hailey-Dunsheath et al. \cite{shd_singletone_2023}. The design incorporates the same lens-coupled aluminum absorber to achieve both a high response and absorption efficiency, and similar niobium interdigitated capacitors (IDCs) with $2\;\mu\mathrm{m}$ tine widths and $10\;\mu\mathrm{m}$ tine spacing to minimize the two-level system noise. The Nb IDCs and coplanar waveguide (CPW) transmission line, to which all 1,008 KIDs are capacitively coupled, are lithographically patterned using a unit cell approach \cite{janssen_inprep_2023}. The unit cell design, as shown in Fig.~\ref{fig:layout} (b), contains 16 (4 by 4) KIDs, each of which consists of the aforementioned aluminuminductor (green elements in Fig.~\ref{fig:layout} (b)), one or two main IDCs setting the pixel’s resonance frequency, a coplanar plate capacitor near the CPW feedline, and an IDC grounding capacitor. The length of the coupling capacitor is matched to the combined width of the main IDCs to maintain an (approximately) constant capacitance ratio and an expected $Q_c \approx30,000$. The IDC capacitor to ground provides approximately 10 times the capacitance of the coupling capacitor to the CPW readout line, thereby minimizing its effect on $Q_c$.
    \par To generate unique resonator frequencies for each detector, the lithographic mask for the Nb components (red elements in Fig.~\ref{fig:layout}(b)) is split in two: (1) a mask containing the top half of each of the main IDCs and the horizontal bar connecting the tines (Fig.~\ref{fig:layout} (b) shaded regions). (2) a mask containing all other Nb components including the bottom half of the main IDCs, coupling capacitor, grounding IDC, two sections of CPW transmission line (below rows 1 and 3) and ground plane (forming a box around each resonator). The combined number of tines in the main IDCs of each KID in the unit cell is varied between 98  and 27 to create 16 resonances uniformly spaced in $\log(\mathrm{frequency})$. The largest capacitor in the array (lowest frequency resonator) is limited by the 900 $\mu\mathrm{m}$ hex-packed pitch, which limits the area available for the main IDCs. This is partially mitigated by allowing low frequency resonators to use IDC space unused by resonators with a smaller IDC (higher frequency). For each consecutive instance of the unit cell the length of all IDC tines is reduced by an integer number of $dL = 0.7\;\mu\mathrm{m}$ through the relative position of the two masks. A few examples of this are indicated in Fig.~\ref{fig:layout} (a). This creates an array with a uniform  design spacing of $df/f \sim 6\times10^{-4}$ and a total design bandwidth of 1.1-2.4 GHz for a 10 nH inductance.
    \begin{figure}[h]
        \centering
        \includegraphics[width=4.76in]{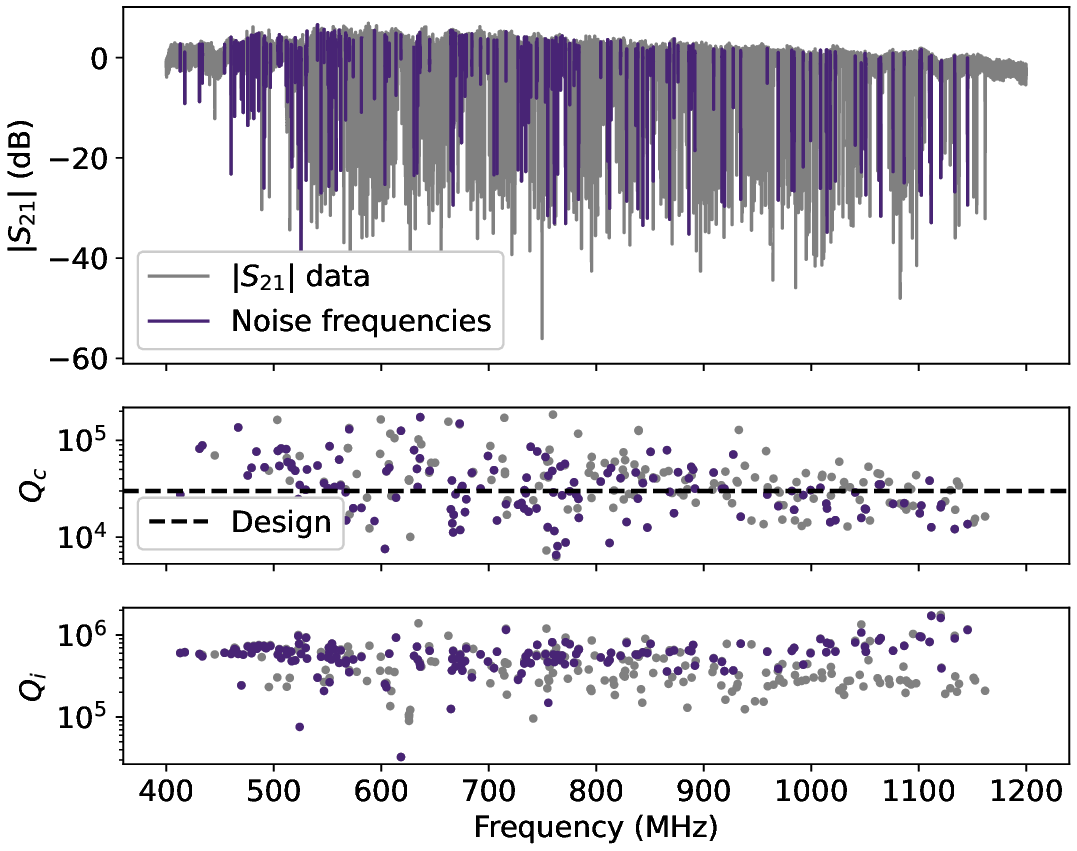}
        \caption{$|S_{21}|$, $Q_c$, and $Q_i$ vs resonance frequency. Purple data corresponds to detectors at which noise was taken for section \ref{subsec:noise}. Resonances were not found below 400 MHz or above 1,200 MHz.}
        \label{fig:vna_rough}
    \end{figure}

\section{Long-wavelength array multitone measurements}
\label{sec:meas}
    We measured a 1,008 pixel array prototype without microlenses in a {BlueFors} dilution refrigerator. {The array is cooled to a temperature of 125 mK for the measurements presented in this paper}. {Without lenses, radiation enters the wafer through the backside with a circular aperture for each pixel defined by holes in the Ti grid.} The readout circuit and array housing are designed to reduce stray electrical power and light. This experimental setup was previously used to demonstrate NEPs down to $2\times10^{-20}\;\mathrm{W}/\sqrt{\mathrm{Hz}}$ \cite{echternach_large_2022}. A detailed description of the experimental setup and single-tone measurements of precursor devices to those presented here can be found in Hailey-Dunsheath et al. \cite{shd_singletone_2023}. For the multitone measurements in this paper, we use a Xilinx Radio Frequency System on a Chip (RFSoC) with firmware \cite{adriankaisinclair_adriankaisinclairprimecam_gateware_design_2023} and software \cite{burgoyne_primecam_readout_2023} developed for the Prime-Cam instrument \cite{sinclair_ccat-prime_2022}. Noise ($S_{xx}$) and response ($dx/dP_{\mathrm{abs}}$) measurements were performed on our array sequentially using the multitone system and the single-tone system described in Hailey-Dunsheath et al. \cite{shd_singletone_2023}. These measurements were consistent between the single-tone and multitone readout systems.

\subsection{Electrical characterization}
    \par A measurement of $S_{21}$ vs readout frequency is presented in Fig.~\ref{fig:vna_rough}. The array yielded 941 out of 1,008 resonances, corresponding to a fabrication yield of 93\%. The resonance frequencies are spread out over a bandwidth of 800 MHz centered on 800 MHz. {No resonances were found below 400 MHz or above 1200 MHz within the measurement band of 10-5,000 MHz}. During design we assumed a 10 nH inductor. However, based on the measurements presented here, as well as measurements of arrays using the same Nb layout in combination with different absorbers, we conclude that the inductors have an inductance of approximately 20 nH. This reduces the resonance frequencies by $\sqrt{2}$ from the design and matches the measured frequencies of the array.
    \par Quality factors ({defined in Zmuidzinas 2012 \cite{zmuidzinas_superconducting_2012}}) are derived from fitting the measured resonance features {at the base blackbody temperature of 5.4 K} and plotted in Fig.~\ref{fig:vna_rough}. We measure a mean internal quality factor of $Q_i=7,300,000$ {with a standard deviation of 200,000}. Hence, our resonator quality factor, $Q_r$ is dominated by the coupling quality factor, for which we measure a mean value of $Q_c = 37,000$ {with a standard deviation of 30,000}. This matches the design value of $Q_c = 30,000$ and the observed spread is within fabrication tolerances.
    \begin{figure}[h]
        \centering
        \includegraphics[width=4.76in]{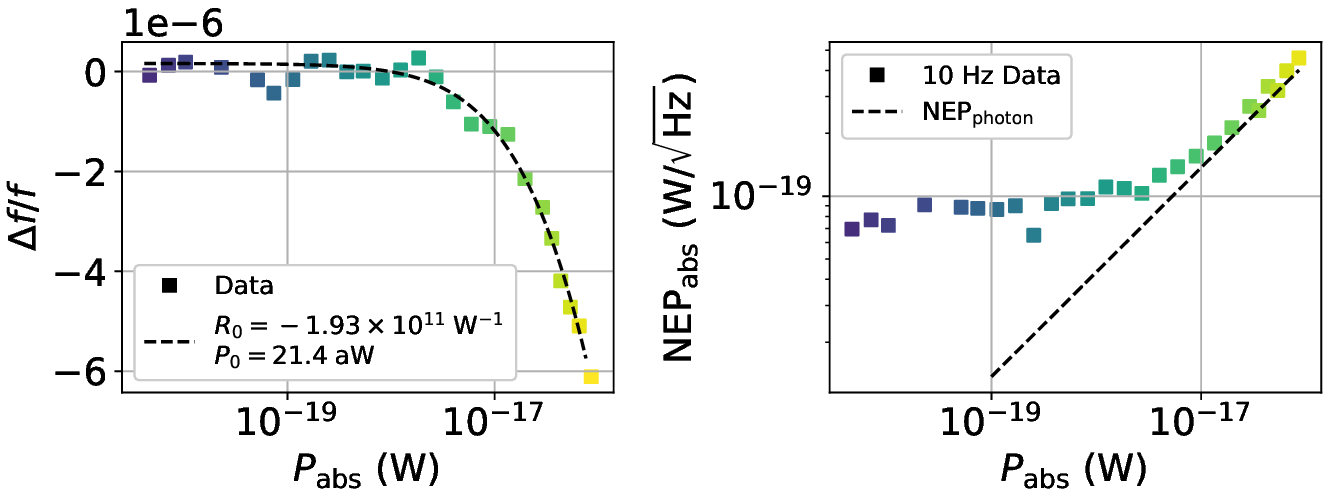}
        \caption{Left: $\Delta f/f$ vs absorbed power with fit for responsivity for a single resonator. Right: $\mathrm{NEP}(10\;\mathrm{Hz})$ vs power. The dashed line indicates the photon noise component. The colorscale corresponds to the absorbed power and matches that used in figure~\ref{fig:nep_hist} (left). }
        \label{fig:fit_example}
    \end{figure}

\subsection{Noise measurements}
\label{subsec:noise}
    \par We measured noise for most pixels in our array, however, during this initial measurement we acquire reliable noise data for only 171 pixels (purple dots in Fig.~\ref{fig:vna_rough}) due to one of three reasons: (1) {nearby resonance frequencies}, (2) off-resonance tone placement, or (3) incorrect readout power. We conservatively rejected 266 out of 941 resonators with a spacing closer than 200 kHz to reject any collided resonators and to ensure a robust operation of our analysis code. Another 201 resonators were rejected for misplaced tone frequencies and 303 resonators for a readout power that was not optimized. The large number of detectors in the latter two categories are the result of our automated method to choose the tone frequency and readout power. To achieve noise measurements with clearance above the amplifier noise, each detector must be biased on-resonance and with a bias power near but below bifurcation. Due to the relatively high amplifier noise temperature in our setup, the tone placement and power optimization constraints were more stringent than expected for PRIMA. Our algorithm that selects readout tone frequencies and powers based on initial $S_{21}$ scans is still under development and only managed to correctly identify these properties for 171 resonators. Optimization of the tone power is particularly difficult: overestimation of the power leads to bifurcation and underestimation of the power leads to amplifier-dominated noise, which are both not representative of the optimized KID noise. PRIMA will have lower amplifier noise than our experimental setup, so power optimization will be less constrained. No resonances were eliminated from the sample based on noise, so this sample is still a representative characterization of the properties of the full array. 
    \begin{figure}[h]
        \centering
        \includegraphics[width=4.76in]{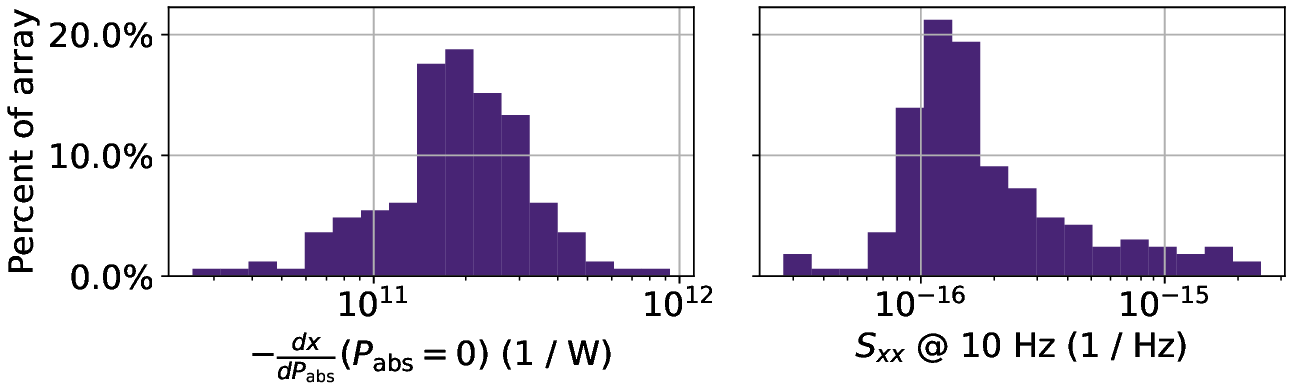}
        \caption{Left: Histogram of responsivity in the limit of no loading. Right: Histogram of $S_{xx}(10\;\mathrm{Hz})$ at the lowest blackbody power, {corresponding to a blackbody temperature of 5.4 K.}}
        \label{fig:resp_hist}
    \end{figure}
    \par To determine the NEP for each of the 171 resonators, we obtain a frequency sweep of $S_{21}$ and a 200 s on-resonance timestream as a function of blackbody temperature (5.4 - 30 K). We estimate the incident power on our detectors from the blackbody temperature, filter transmission and setup geometry. The true power absorbed in our detectors, $P_{\mathrm{abs}}$, is then determined from the photon shot noise at high blackbody temperatures, following the methodology of Janssen et al. \cite{Janssen2013}. 
    \par A nonlinear resonance model is fit to the measured resonance feature to determine the resonance frequency and fractional frequency shift $x=\Delta f/f$ with respect to the lowest blackbody temperature, $T_{bb} = 5.4\;\mathrm{K}$. Fig.~\ref{fig:fit_example} shows the results of this measurement for one example resonator. To determine the responsivity of the resonator, the $x(P_{\mathrm{abs}})$ data is fitted to the integrated form of the Mattis-Bardeen responsivity model, given by $\frac{dx}{dP_{\mathrm{abs}}}=R_0\left(1 + P_\mathrm{abs}/P_0\right)^{-1/2}$. Here $R_0=\frac{dx}{dP_{\mathrm{abs}}}(P_{\mathrm{abs}}=0)$ is the responsivity in the limit of no loading and $P_0$ is a parameter characterizing the power roll off. A histogram of $R_0$ for all 171 resonators is shown in Fig.~\ref{fig:resp_hist} (left). 
    \par
\begin{figure}[h]
        \centering
        \includegraphics[width=4.76in]{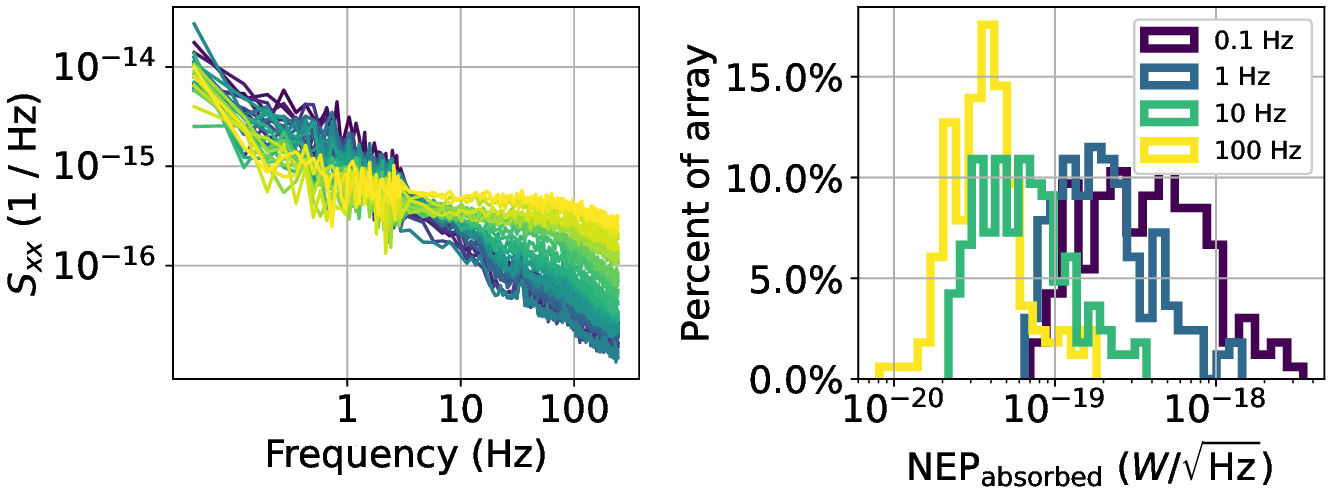}
        \caption{Left: $S_{xx}$ versus noise frequency for a single KID. The colorscale corresponds to the absorbed power, as in figure~\ref{fig:fit_example}. Right: Histograms of NEPs at 0.1, 1, 10, and 100 Hz noise frequency. 73\% of the KIDs have an NEP of less than $1\times10^{-19}\;\mathrm{W}/\sqrt{\mathrm{Hz}}$ at 10 Hz noise frequency.}
        \label{fig:nep_hist}
    \end{figure}

 Fig.~\ref{fig:resp_hist} (right) shows a histogram of the measured fractional frequency noise at a noise frequency of 10 Hz, $S_{xx}(10\;\mathrm{Hz})$, at the lowest blackbody power. Fig.~\ref{fig:nep_hist} (left) shows $S_{xx}$ as a function of optical loading.  The phase timestream data is converted to a fractional frequency shift using a polynomial fit to the $S_{21}$ data. The $S_{xx}$ spectra have a white component from photon noise that rises with blackbody power. 1/f noise from two-level systems is present, and limits the noise at the lowest blackbody power.
    \par The NEP is given by $\mathrm{NEP}=\sqrt{S_{xx}}/\frac{dx}{dP_{\mathrm{absorbed}}}\sqrt{1+\omega^2\tau^2}$, where $\frac{\omega}{2\pi}$ is the noise frequency of the noise and $\tau$ is the detector response time, which is limited by the quasi-particle lifetime, $\tau_{qp}$, for our KIDs. Fig.~\ref{fig:fit_example} (right) shows an example of the measured NEP as a function of absorbed power. This data is well described by $\mathrm{NEP}_{\mathrm{detector}}^2 = \mathrm{NEP}_{\mathrm{photon}}^2 + \mathrm{NEP}_\mathrm{recombination}^2 + \mathrm{NEP}_{0}^2$. Here $\mathrm{NEP}_{\mathrm{photon}} = \sqrt{2h\nu (1 + n_\gamma)P_{\mathrm{abs}}}$ is the NEP set by the photon shot noise (dashed line), $\mathrm{NEP}_\mathrm{recombination}=\sqrt{4\Delta_0^2\Gamma_r/\eta_{pb}^2}$ is the recombination NEP, and $\mathrm{NEP}_0$ is the remaining power-independent NEP. {$\mathrm{NEP}_0$ is dominated by quasiparticle generation-recombination noise and two-level system noise in the interdigitated capacitors.} $h\nu$ is the photon energy, $n_\gamma$ is the photon occupation number in the detector, $\Delta_0=1.76k_BT_c$ is the superconducting gap energy, $\Gamma_r$ is the recombination rate \cite{zmuidzinas_superconducting_2012}, and $\eta_{pb}$ is the pair-breaking efficiency. We approximate $\eta_{pb}$ to be 0.5 \cite{shd_singletone_2023}, and we calculate that $n_\gamma$ is negligible within our error bars.
    \par Fig.~\ref{fig:nep_hist}(right) shows the histogram of the detector limited $\mathrm{NEP}$ {at a blackbody temperature of 5.4 K} at multiple noise frequencies. {The mean NEP at 10 Hz is $8.2\times10^{-20}\;\mathrm{W}/\sqrt{\mathrm{Hz}}$ with a standard deviation of $7\times10^{-20}\;\mathrm{W}/\sqrt{\mathrm{Hz}}$.} 73\% of the pixels have an $\mathrm{NEP}(10\;\mathrm{Hz})$ below $1\times10^{-19}\;W/\sqrt{\mathrm{Hz}}$. This percentage is significantly improved at 100 Hz. The roll off of the responsivity is calculated using a time constant of $\tau=930$ $\mu$s, measured using the single-tone setup for the resonator at 1101.01 MHz, to get an estimate of the NEP histogram at higher noise frequencies. This roll off is only significant for the 100 Hz data. 

\section{Summary and Discussion}
\label{sec:summary}
    We present the design, initial characterization and array-level performance statistics of a flight-like long-wavelength PRIMA array. We find 941 resonances, indicating a fabrication yield of 93\%. The readout bandwidth and quality factors are consistent with the design and match the instrument requirements. Using an RFSoC-based multitone readout system we measured the noise and response for a subset of 171 KIDs. We measured an $\mathrm{NEP}(10\;\mathrm{Hz})$ of less than $1\times10^{-19}\;\mathrm{W}/\sqrt{\mathrm{Hz}}$ for 73\% of the KIDs studied, and confirmed that our multitone and single-tone measurements are consistent. We observe an improvement in the NEP at higher noise frequencies as a result of TLS noise, which is consistent with most KID devices.  Modulating signals at higher frequencies offers a path to higher sensitivities if needed, a capability readily available for PRIMA with its cryogenic steering mirrors. More broadly, improvements in TLS translate directly to improvements in sensitivity. Our parallel plate capacitors used in our short-wavelength devices show a reduction in $S_{xx}(10\;\mathrm{Hz})$ of 30\% with respect to our IDC capacitors, and thereby provide a pathway to improving the NEP. 
    \par We expect to recover many of the closely spaced resonators through improvements to the software, using a chip with a wider resonance frequency spacing, and trimming capacitors to separate overlapping resonators \cite{liu_superconducting_2017}. 
    {Other planned work includes: 1) optimization of readout tone frequency and power tuning at array level to provide access to more pixels in the array, and 2) LED mapping and verifying low inter-pixel cross talk with the multitone readout, similar to the work of Liu et al. in the TIM balloon arrays \cite{liu_ltd2023}. }

\begin{acknowledgements}
The research was carried out at the Jet Propulsion Laboratory, California Institute of Technology, under a contract with the National Aeronautics and Space Administration (80NM0018D0004). This work was funded by the NASA (Award No. 141108.04.02.01.36)—to Dr. C. M. Bradford. We would like to thank Adrian Sinclair for teaching us how to operate the Prime-Cam RFSoC and adding software features for us. We would also like to thank Chris Groppi's group at Arizona State University for providing us with a housing for our RFSoC. 
\end{acknowledgements}

\bibliographystyle{sn-mathphys}
\bibliography{bibliography.bib}

\end{document}